\documentclass[aps,prl,twocolumn,footinbib,letterpaper,preprintnumbers,showpacs]{revtex4}
\pdfoutput=1
\usepackage{amsfonts}
\usepackage{amsmath}
\usepackage{amssymb}
\usepackage{amsthm}
\usepackage{graphicx}
\usepackage{color}

\begin{document}

\title{$\boldsymbol{b-\tau}$ Yukawa (Non-)Unification in the CMSSM}

\author{Maurizio Monaco}
\email{mmonaco@sissa.it}
\author{Martin Spinrath}
\email{spinrath@sissa.it}
\affiliation{SISSA/ISAS and INFN,\\
Via Bonomea 265, I-34136 Trieste, Italy}

\preprint{SISSA 33/2011/EP}
\pacs{12.10.Kt, 12.60.Jv}

\begin{abstract}
\noindent
Supersymmetric Grand Unification usually provides unification of the bottom quark and the tau lepton Yukawa couplings at the GUT scale. In the CMSSM this can be realised only for a very particular choice of parameters.
In this letter we study the GUT scale ratio $y_\tau/y_b$ for less peculiar parameters in the large $\tan \beta$ regime and identify one parameter region preferred by current experimental data. In this region, which is well within the reach of the LHC, the ratio is very close to the recently proposed value of 3/2.
\end{abstract}

\maketitle

\section{Introduction}

Supersymmetric (SUSY) Grand Unified Theories (GUTs) are prompted as one of the most appealing extensions of the Standard Model (SM). Among other things they offer the opportunity to relate the flavour sector of the quarks and the leptons to each other. The most praised example for this is the fact that in
many GUT models the Yukawa couplings of the bottom quark and the tau lepton unify at the GUT scale, as it was at first obtained in \cite{Georgi:1974sy}.
Under the assumption that no new physics appears between the electroweak and the GUT scale, the mass splitting observed at low energy would then be entirely owed to renormalization group running.
In fact, if in a bottom-up approach the Yukawa couplings are evolved under the above assumptions up to the GUT scale, they do not unify, being their ratio $y_\tau/y_b$ roughly 1.3 for $\tan \beta$ between 5 and 50. Yukawa unification could be recovered by finite SUSY threshold loop corrections \cite{SUSYthresholds}. Nevertheless, within the Constrained Minimal Supersymmetric Standard Model (CMSSM) the SUSY breaking parameters would in this case have to fulfill a very peculiar pattern, for recent discussions see, e.g.\ \cite{bTauUnification}. In most of the CMSSM parameter space the ratio is actually shifted to larger values.

In recent papers, one of the authors \cite{Antusch:2008tf, Antusch:2009gu} realised that there is an alternative to $b-\tau$ Yukawa unification. If the Yukawa couplings are generated effectively by a dimension five operator with an additional adjoint $SU(5)$ Higgs representation, the $y_\tau/y_b$ ratio would be predicted to be $3/2$, easily achievable within the CMSSM for values of $\tan \beta \gtrsim 20$.

In this letter we go beyond these previous studies. First, we show the ratio $y_\tau/y_b$ at the GUT scale in the $m_0$-$M_{1/2}$-plane for different values of $\tan \beta$ and $a_0 = A_0/m_0$. This is a useful tool for model builders, if their models predict either a certain SUSY breaking scheme close to the CMSSM or a certain GUT scale value for $y_\tau/y_b$. For an illustrative discussion of the interplay between the SUSY spectrum and a non-standard $b$-$\tau$ Yukawa coupling ratio, see, for instance, \cite{Antusch:2011sq}.
Second, we provide not only the GUT scale ratio $y_\tau/y_b$, but also some up-to-date experimental constraints within these planes, to identify preferred parameter regions and values for $y_\tau/y_b$.

\section{Experimental Constraints}

To calculate the SUSY spectrum we use {\tt SOFTSUSY} \cite{Allanach:2001kg} and
focus on the most important constraints on the CMSSM for large $\tan \beta$. We collect the numerical values of the constraints as well as the most important SM input parameters in Tab.\ \ref{tab:Input}.

\begin{figure*}
\centering
\includegraphics[scale=0.235]{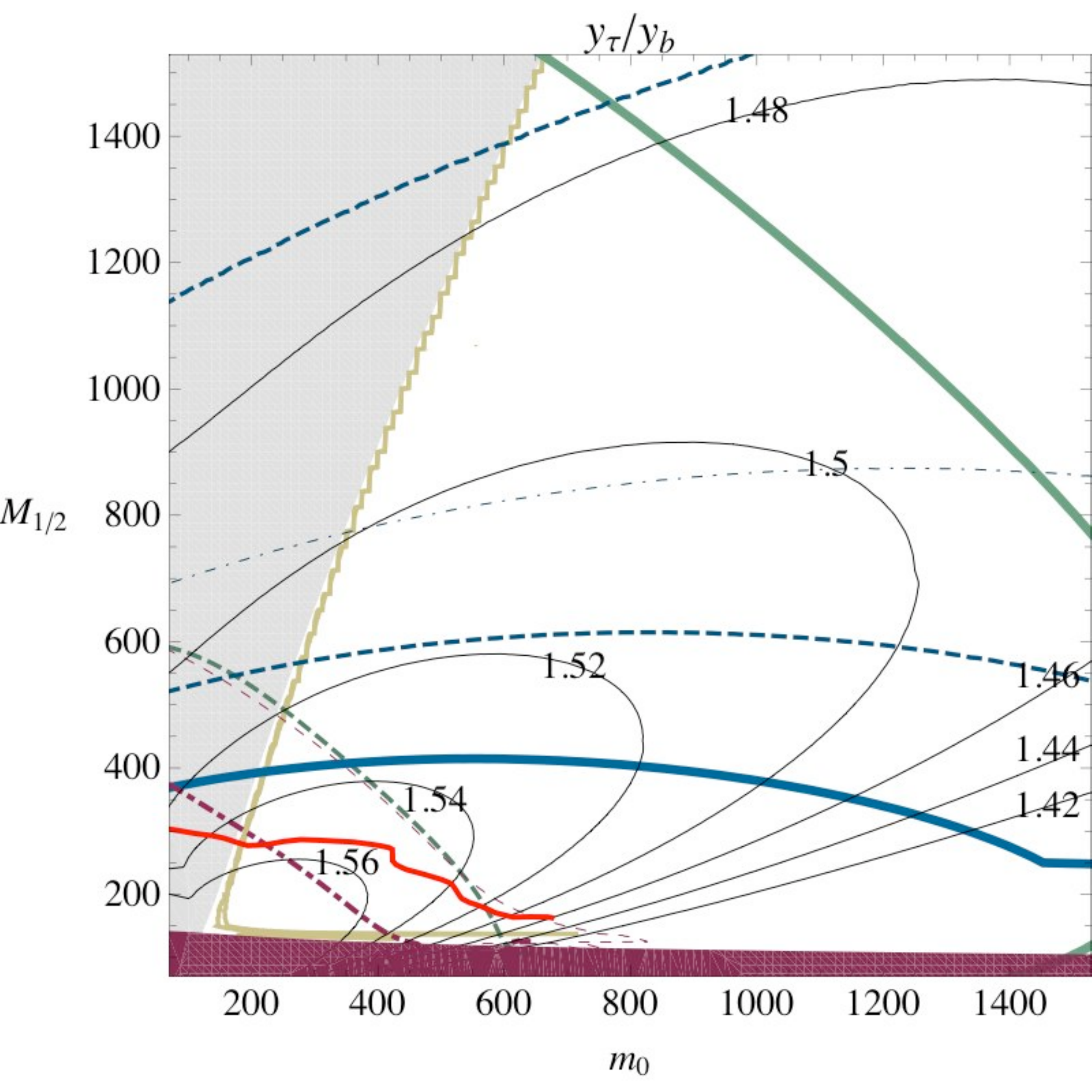}  \quad
\includegraphics[scale=0.235]{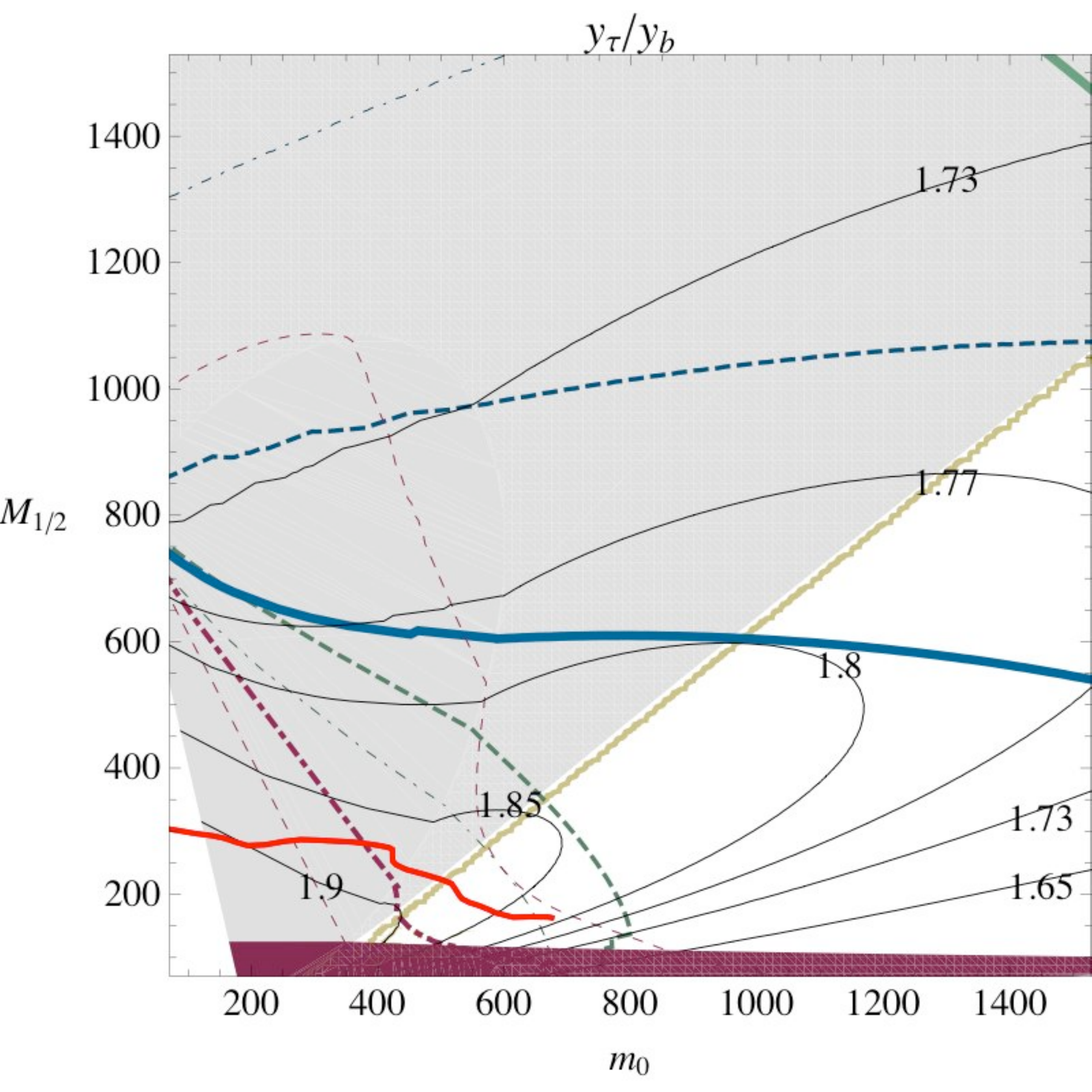}
\newline
\includegraphics[scale=0.235]{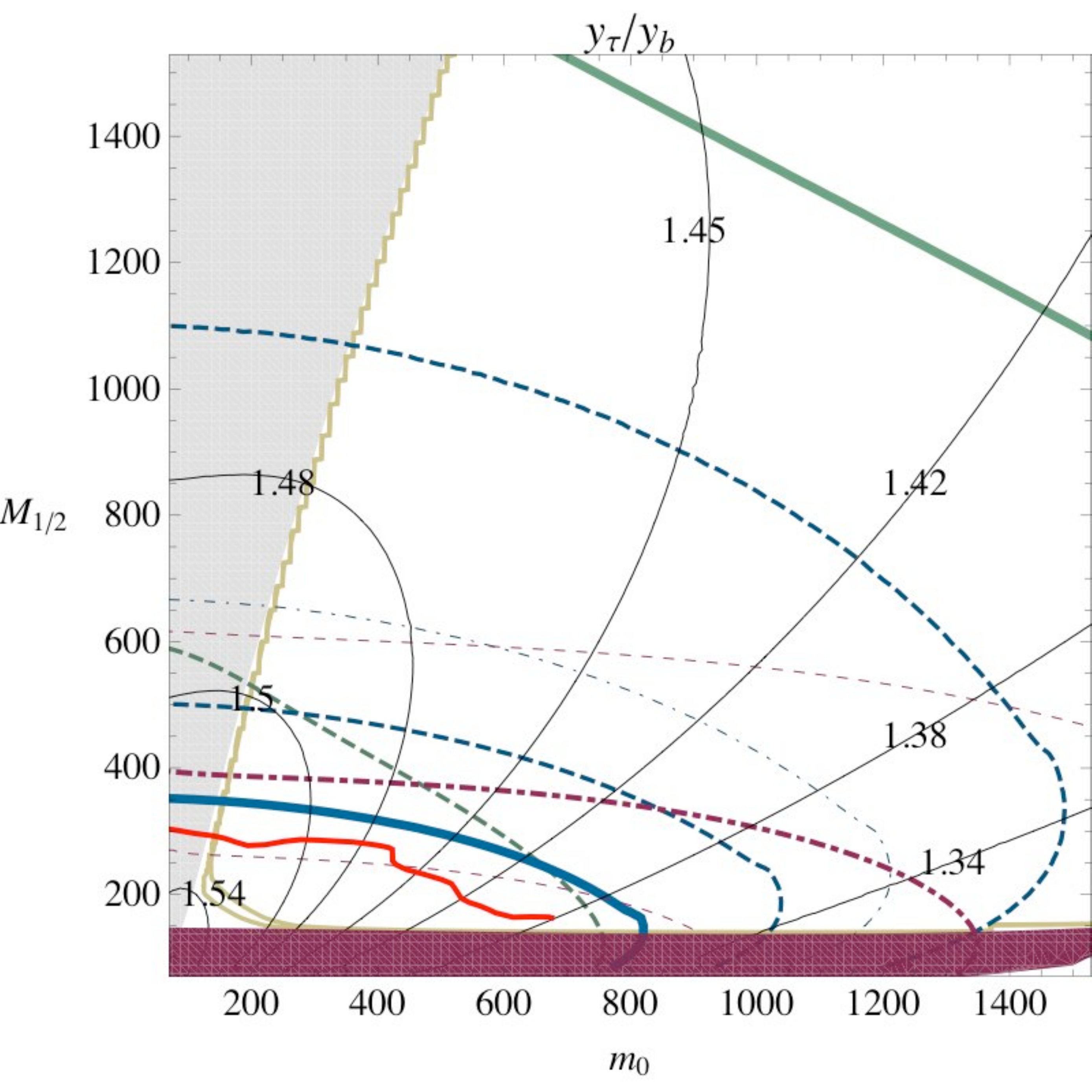}  \quad
\includegraphics[scale=0.235]{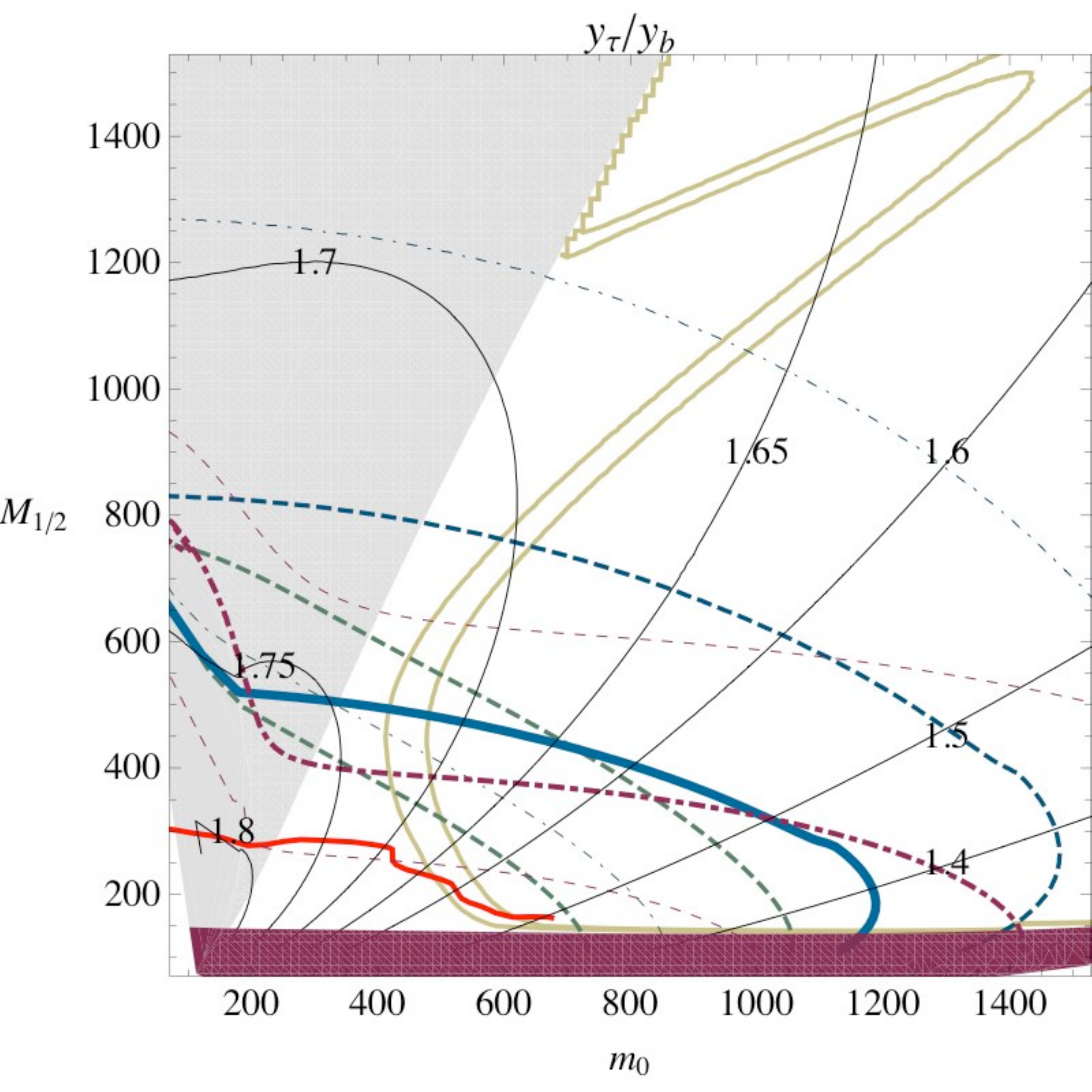}
\newline
\includegraphics[scale=0.235]{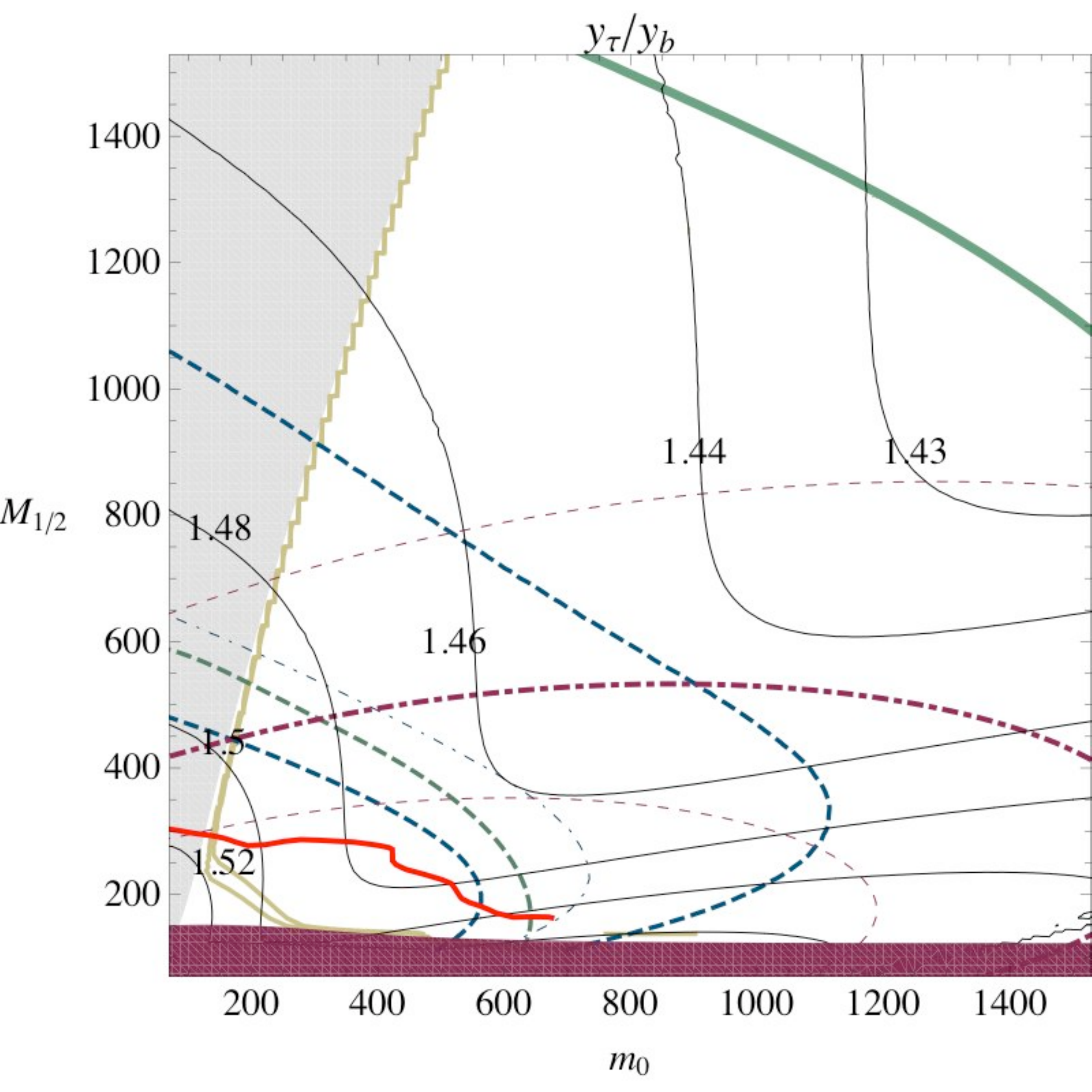}  \quad
\includegraphics[scale=0.235]{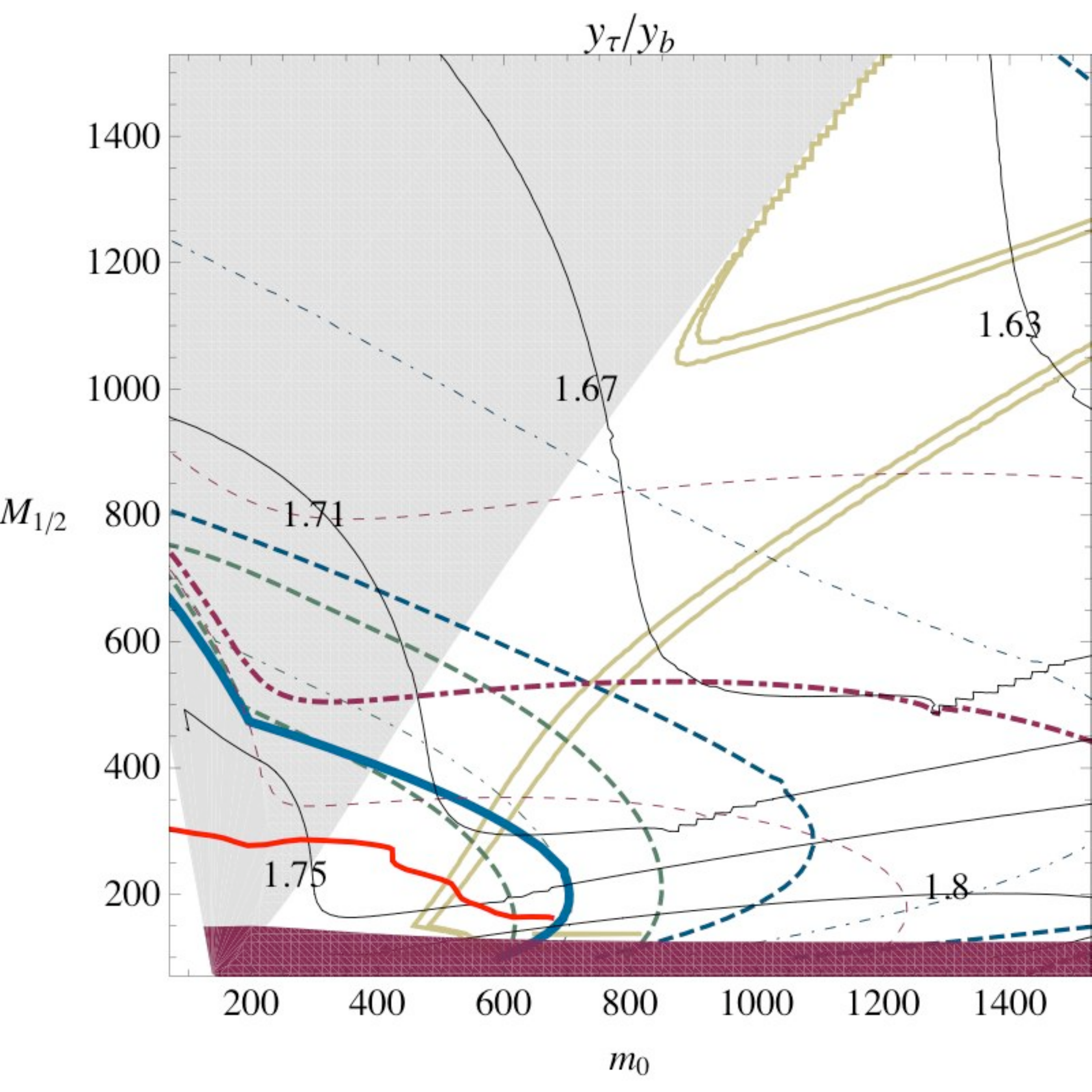}
\newline
\caption{Dependence of $y_\tau/y_b$ at the GUT scale on $m_0$ and $M_{1/2}$ for $\tan \beta = 30$ (left column) and $\tan \beta = 50$ (right column), $a_0 = -2$, 0, and 2 (from top to bottom). Contours of constant $y_\tau/y_b$ are given by the black lines. The best-fit, 1$\sigma$, and 3$\sigma$ contours for $BR(b\to s\gamma)$ are given by the thin dot-dashed, thick dashed, and thick straight blue line. The best-fit, 1$\sigma$, and 3$\sigma$ lines for $\delta a_\mu$ are given by the thin dot-dashed, thick dashed, and thick straight green line. The 3 $\sigma$ region for CDM is depicted by the yellow lines and the stau LSP region by the grey region. The LEP limit for the lighter chargino mass is given by the purple region. The Higgs mass is given by the purple lines. The dot-dashed line corresponds to $m_h = 114$~GeV, while the lower (upper) thin dashed one to $m_h = 111 (117)$~GeV. The ATLAS exclusion bound coming from the SUSY search with one lepton, jets and missing transverse energy in the final state is given by the straight red line.
\label{fig:Results}} 
\end{figure*}

In the flavour sector the two most prominent constraints for the large $\tan \beta$ regime are given by the observables $BR(b\to s \gamma)$ and $BR(B_s \to \mu \mu)$, that are both $\tan \beta$ enhanced.
The calculation of these observables is performed with {\tt SuperIso} \cite{SuperIso}. Nevertheless, it turned out that only the former gives relevant constraints in the $m_0$-$M_{1/2}$ planes and therefore we present only this one in our final plots in Fig.\ \ref{fig:Results}.

In the SM, there is an approximate 3$\sigma$ tension between the experimentally determined anomalous magnetic moment of the muon and its theoretical prediction, $\delta a_\mu$ \cite{Hagiwara:2011af}.
Within SUSY models this tension can be relaxed by finite contributions, that are $\tan \beta$ enhanced as well.
We calculate these ones with the {\tt SuperIso} package \cite{SuperIso}.

We also show the 3$\sigma$ region preferred by the experimental determination of the Cold Dark Matter (CDM) relic density as provided by the WMAP collaboration \cite{Larson:2010gs}, which we calculate with {\tt DarkSusy} including Sommerfeld enhancement \cite{DarkSusy}.
This translates into a bound on the parameter space under the assumption that we have standard cosmology, R-parity is conserved, and dark matter is made of the lightest neutralino.

Furthermore, we include the LEP mass limit for the chargino \cite{Nakamura:2010zzi} and the exclusion bound from the ATLAS SUSY search with one lepton, jets and missing transverse energy in the final state \cite{ATLAS_line}. The other ATLAS \cite{ATLAS} and CMS results \cite{CMS} are similar or even weaker.
The ATLAS bound is given only for a specific value of $A_0$ and $\tan \beta$, but as discussed, e.g.\ in \cite{Allanach:2011ut}, it depends only weakly on these two parameters.

\begin{table}
\caption{Numerical values for the experimental input used in our scans. We only list third generation fermion masses and the value for $\alpha_s$, since the other SM parameters have a negligible influence on the GUT scale $y_\tau/y_b$. The errors for $BR(b\to s \gamma)$ are added linearly.  \label{tab:Input}}
\begin{ruledtabular}
\begin{tabular}{lc}
$m_b(m_b)$ & 4.2 GeV \\
$m_\tau^{\text{pole}}$ & 1.777 GeV \\
$m_t^{\text{pole}}$ & 173.3 GeV \\
$\alpha_s(M_Z) $ & 0.1176 \\ \hline
$m_{\tilde{\chi}_1^\pm}$ & $ > 94$ GeV (CL 95\%) \cite{Nakamura:2010zzi}    \\
$10^{10} \delta a_\mu$  & $26.1 \pm 8.0$ \cite{Hagiwara:2011af}  \\
$10^4 BR(b\to s \gamma)$ &  $3.61 \pm 0.18$ \cite{Nakamura:2010zzi}  \\
$\Omega_{\text{CDM}} h^2$ & $0.1120 \pm 0.0056$  \cite{Larson:2010gs}   \\
\end{tabular}
\end{ruledtabular}
\end{table}

\section{Results}

In Fig.\ \ref{fig:Results} we present our results in terms of plots of the $m_0$-$M_{1/2}$ plane in the CMSSM for $\tan \beta = 30$ and $50$ and $a_0 \equiv A_0/m_0 = -2$, $0$, and $2$. For smaller $\tan \beta$ the GUT scale ratio $y_\tau/y_b$ is approximately constant due to the lack of enhancement of the SUSY threshold corrections. The interesting regime begins with $\tan \beta \gtrsim 25$ since there the influence of the SUSY thresholds becomes comparable to the experimental error on the $b$-quark mass.

The dependence of $y_\tau/y_b$ in terms of the CMSSM parameters shows a peculiar behaviour.
For large (positive) $a_0$ and $M_{1/2} \lesssim m_0$ the SUSY threshold contributions from gauge interactions and the ones proportional to the trilinear couplings add up (within our sign convention) leading to an almost linear shape of the $y_\tau/y_b$ contour lines in the $m_0$-$M_{1/2}$ plane.
For lower values of $a_0$, and also for $M_{1/2} \gtrsim m_0$ the two SUSY threshold contributions
partially cancel leading to curvy $y_\tau/y_b$ contours.

As expected we do not find anywhere values near $y_\tau/y_b = 1$. Within the CMSSM this happens only for extreme cases \cite{bTauUnification}, far outside the shown regions, where the contributions to the SUSY threshold corrections proportional to the trilinear coupling are dominant over the gauge contributions, especially the SUSY QCD one.

We find, as already known, that $y_\tau/y_b$ at the GUT scale is larger than 1. For $\tan \beta =30$ this ratio is roughly between 1.3 and 1.6 and for $\tan \beta = 50$ it grows up to values between roughly 1.4 and 1.9, in line with \cite{Antusch:2009gu}.

Now we can easily read off from the plots the experimentally preferred regions.
By looking at the CDM 3$\sigma$ strip, which is the most stringent constraint, we see that
for $\tan \beta = 30$ only the stau coannihilation region is allowed. Interestingly, for $a_0 = 0$ there is a small region around $m_0 \approx 200$~GeV and $M_{1/2} \approx 525$~GeV, see Tab.~\ref{tab:GoodPoint}, where the relic density lies within the WMAP bound, the predictions for $BR(b\to s \gamma)$ and $\delta a_\mu$ are in agreement with experiments with less than 1$\sigma$.
Note that there $y_\tau/y_b$ is very close to 3/2, a relation suggested by one of us in \cite{Antusch:2009gu}. For the other two values of $a_0$ shown we do not find a similarly good region, since the tension between $\delta a_\mu$ and $BR(b\to s \gamma)$ grows. Nevertheless, the ratio $y_\tau/y_b$ is still close to $3/2$.

\begin{table}
\caption{Numerical values of the input parameters for the point, which is in best agreement with the experimental constraints. We give here also the light Higgs mass, some important SUSY particle masses and values for some relevant observables. \label{tab:GoodPoint}}
\begin{ruledtabular}
\begin{tabular}{lc}
$m_0$ & 200 GeV \\
$M_{1/2}$ & 525 GeV \\
$a_0$ & 0 \\
$\tan \beta$ & 30 \\
$\text{sgn}(\mu)$ & +1 \\ \hline
$m_{h}$ & 116 GeV \\
$m_{\tilde{g}}$ & 1.2 TeV \\
$m_{\tilde{q}}$ & 1.1 TeV \\
$m_{\tilde{\chi}^0_1}$ & 216 GeV \\
$m_{\tilde{\chi}^{\pm}_1}$ & 409 GeV \\
$m_{\tilde{\tau}_1}$ & 224 GeV \\
$m_{\tilde{t}_1}$ & 853 GeV \\
$m_{\tilde{b}_1}$ & 990 GeV \\
 \hline
$10^{10} \delta a_\mu$  & $18.4$ \\
$10^4 BR(b\to s \gamma)$ &  $3.48$  \\
$10^9 BR(B_s \to \mu \mu)$ & $3.49$ \\
$\Omega_{\text{CDM}} h^2$ & $0.119$ \\
\end{tabular}
\end{ruledtabular}
\end{table}

For $\tan \beta = 50$ the situation changes. First of all, the disfavoured region, where the stau is the lightest SUSY particle, becomes quite large.
For $a_0 = -2$ there is no region in which both the tension for $\delta a_\mu$ and that for $BR(b\to s \gamma)$ are less than 2$\sigma$ while simultaneously fulfilling the CDM bound.
For the other two $a_0$ cases one clearly sees that $\delta a_\mu$ prefers a rather light spectrum, while $BR(b \to s \gamma)$ prefers a heavier spectrum. In combination with the CDM and Higgs mass bounds in the case $a_0 = 2$ the light mass region is nevertheless disfavoured, while in the case $a_0 = 0$ $BR(b \to s \gamma)$ excludes this region. For the heavier regions, despite not curing the SM tension with $\delta a_\mu$, we find a preferred $y_\tau/y_b \approx 1.65$.

The collider constraints from ATLAS \cite{ATLAS_line, ATLAS} and CMS \cite{CMS} are not important so far.
We show here only the ATLAS exclusion bound coming from the SUSY search with one lepton, jets and missing transverse energy in the final state, but the other ones are similar or worse.
We have checked that our preferred point as given in Tab.\ \ref{tab:GoodPoint} is not excluded yet.
This might change in the near future. The LHC has already taken more than 1~fb$^{-1}$ of data, while the exclusion bounds discussed here are derived from only 35~pb$^{-1}$ of data. Therefore, in the next year most of the preferred regions discussed here will be tested.

\section{Summary and Conclusions}

For realistic flavour models embedded into a SUSY GUT it is essential to know the GUT scale Yukawa coupling ratios.
In this letter we present the most prominent and generic one, the ratio $y_\tau/y_b$, for different slices of the CMSSM parameter space. The variation of this ratio in terms of the SUSY parameters is enhanced by the SUSY threshold corrections which are very important in the large $\tan \beta$ regime. To determine preferred regions in these planes we also show various constraints from different experiments. The strongest one comes from the WMAP measurement of the relic density. Together with inputs from flavour physics ($BR(b \to s \gamma)$) and from measurements and calculations of the anomalous magnetic moment of the muon, we find a region with small tension to all observables, where $y_\tau/y_b$ at the GUT scale is very close to 3/2. This happens for a relatively light spectrum, with the first two generation squark and gluino masses around 1.2~TeV, see Tab.\ \ref{tab:GoodPoint}, which will be tested in the very near future by the LHC.

\section{Acknowledgements}

We would like to thank Stefan Antusch and Andrea Romanino for useful discussions.
This work is partially supported by the Italian government under the project number PRIN 2008XM9HLM ``Fundamental interactions in view of the Large Hadron Collider and of astro-particle physics''.

\end{document}